\documentclass[%
 reprint,
 amsmath,amssymb,
aps,
pra,
]{revtex4-2}

\usepackage{graphicx}
\usepackage{dcolumn}
\usepackage{bm}
\usepackage{hyperref}
\usepackage{xcolor}
\usepackage{soul}

\begin{document}

\preprint{APS/123-QED}

\title{A comparative study of focusing with scalar and vector beams in an active Raman gain system}

\author{Partha Das$^{}$}
\email{partha.2015@iitg.ac.in}
\author{Tarak Nath Dey$^{}$}
\email{tarak.dey@iitg.ac.in}
\address{Department of Physics, Indian Institute of Technology Guwahati, Guwahati 781039, Assam, India}


\begin{abstract}

We investigate the focusing characteristics of scalar and vector beams within an atomic medium. An active-Raman-gain configuration is employed to achieve significant Kerr nonlinearity in a four-state atomic system. The probe beams can attain focusing within the medium through careful selection of input beam intensities and the spatial profile of the control field. We analytically derive the linear and third-order nonlinear susceptibilities for both scalar and vector probe beams. Our observations indicate that, in addition to the energy transfer from the control beam to the probe beam, the giant cross-Kerr nonlinearity facilitates the focusing of the scalar probe beam into a significantly smaller spot size. Conversely, the vector probe beams exhibit gain-induced narrowing. Furthermore, we evaluate the state of polarization for the vector beam at the minimum beam waist, observing a polarization rotation and a change in ellipticity during propagation. Through the mechanism of focusing, we achieve a reduced spot size for the probe beam, which may have substantial implications for resolution enhancement in microscopy applications.
 
\end{abstract}

\maketitle


\section{\label{sec:level1}INTRODUCTION}
Precise control over light focusing is essential for minimizing diffraction, dispersion, and absorption—key factors that limit performance in image processing, high-resolution imaging, and optical lithography. The dispersive properties can be effectively manipulated by applying a coherent field to an unoccupied transition \cite{intro1,intro2,intron3}. This method of dispersion engineering has yielded significant advancements and has fostered compelling applications in the field of nonlinear optics \cite{introdis1,introdis2}. The phenomenon of self-focusing of light within a nonlinear medium has emerged as a topic of considerable significance \cite{intro3,intro4,intro5}. Typically, the self-focusing of an optical beam is attributed to the nonlinear optical Kerr effect, representing a universal nonlinear phenomenon observed across a wide range of optical domains. Kerr induced self-focusing in plasmas was first proposed in 1962 \cite{intro6}, with its theoretical model soon established \cite{intro7,intro8}. It was later confirmed experimentally by Lallemand $\textit{et al.}$ and analytically explained by Vlasov $\textit{et al.}$ \cite{intro9,intro10,intro11}. In a Kerr nonlinear system, an optical beam can undergo evolution resulting in spikes of infinite amplitude over a finite propagation distance. This phenomenon is referred to as wave collapse. The physical phenomenon of wave collapse has been observed in various contexts such as plasma waves \cite{intro12,intron13}, Bose-Einstein condensates or matter waves \cite{intron14}, capillary-gravity waves in deep water \cite{intron15}, and astrophysics \cite{intron16}. Moreover, self-focusing and filamentation of the optical field have been observed in transparent media as well as in absorptive resonance gases \cite{intron17,intro17}. It is noteworthy that during focusing, a high-power laser beam may experience stimulated Raman scattering (SRS), resulting in the majority of its energy being transferred to a Stokes wave \cite{intro18}. The concept of gain-focusing associated with propagation in water is discussed in further detail in \cite{intro19}.

Kerr nonlinearity, which is crucial for the majority of nonlinear processes, has traditionally been realized in passive optical media, such as glass-based optical fibers \cite{intro20}. In passive optical media, nonlinear effects are relatively weak. Achieving sufficient nonlinear phase shifts requires a long propagation distance or high light intensities. These conditions are essential for the formation of optical solitons. Electromagnetically induced transparency (EIT) has garnered significant attention in highly resonant optical media \cite{intro21}. Quantum coherence and interference, induced by a control field, drastically reduce the absorption of a weak probe field. A strong one-photon resonance tuning mitigates the medium’s opacity. This effect makes an otherwise opaque optical medium transparent \cite{intro18,intro19,intro20}. The propagation of electromagnetic pulse in EIT systems demonstrates both a substantial reduction in group velocity and an enhancement of Kerr nonlinearity. These characteristics have prompted to explore soliton-pair solutions within EIT systems \cite{intron21,intron22}. Additionally, contemporary theoretical investigations have predicted a diverse array of fractional solitons \cite{intron23,intron24,intron25}. The enhancement of Kerr nonlinearity at various transparency frequencies has been examined \cite{intron26,intron27}. Moreover, Mukherjee $\textit{et al.}$ have conducted investigations into the impact of quintic nonlinearity on modulation instability in multiple coupled quantum wells \cite{intron28}. They have also analyzed modulation instability influenced by the relative phase of applied optical fields in the EIT regime \cite{intron29}. Numerous theoretical efforts have been made to exploit the robust interactions of Rydberg atoms for the realization of Kerr nonlinearity \cite{intron30,intron31,intron32,intron33}. In contrast to the absorptive characteristics of the EIT framework, the Active Raman Gain (ARG) approach has attracted substantial theoretical and experimental focus \cite{intro22,intro23,intro24}. The fundamental principle underlying the ARG scheme involves the amplification of the probe field through stimulated absorption facilitated by the control field. This system is operationally effective at ambient temperatures, which reduces both signal attenuation and distortion \cite{intro25}. Significant attention has been devoted to the study of third-order self-Kerr \cite{intron34,intron35,intron36} and cross-Kerr nonlinear responses \cite{intron37,intron38}. The anisotropic Kerr nonlinearity may be utilized in nonlinear optical characterization techniques and applications for light manipulation \cite{intron39,intron40}. Among other nonlinear process, the physical mechanism for nonlinear trapping has been recently reported \cite{intron41,intron42}. Following its successful application in atomic systems, the exploration of nonlinear optical properties has seamlessly progressed into the domain of semiconductor quantum wells and quantum dots \cite{intron43,intron44,intron45}. 

In this study, we present an investigation into the focusing properties of scalar and vector beams (VBs) within an atomic system. The atomic configuration utilized comprises a four-level arrangement that employs an ARG scheme. A comparative analysis is performed on the focusing effects associated with probe scalar beams and VBs. It has been observed that cross-Kerr-induced focusing occurs with scalar beams, while VBs experience gain-induced narrowing. Importantly, under conditions of two-photon resonance, the probe beam can achieve focusing or narrowing within the medium through a careful selection of input beam intensity. To substantiate our findings, we employ the probability-amplitude method for the calculation of both linear and third-order nonlinear susceptibilities of the probe beams. Additionally, we investigate the polarization state of the VB at the point of minimum beam radius. The resultant narrowing of the VB leads to a reduction in spot size, indicating significant potential for applications aimed at enhancing resolution. 

The structure of this paper is organized as follows. Section I provides a concise introduction to the concepts of light focusing, Kerr nonlinearity, their applications, and the findings from our research. The theoretical framework utilized in this study is outlined in Sec. II. Section III details the results obtained from the investigation of  probe scalar beam, supplemented by thorough explanations. In Sec. IV, we conduct a comparative analysis with  vector probe beam, offering clarifying insights throughout. Finally, Sec. V presents the conclusions drawn from our study.
 
\begin{figure}[ht!]
	\centering
	\includegraphics[width=\linewidth]{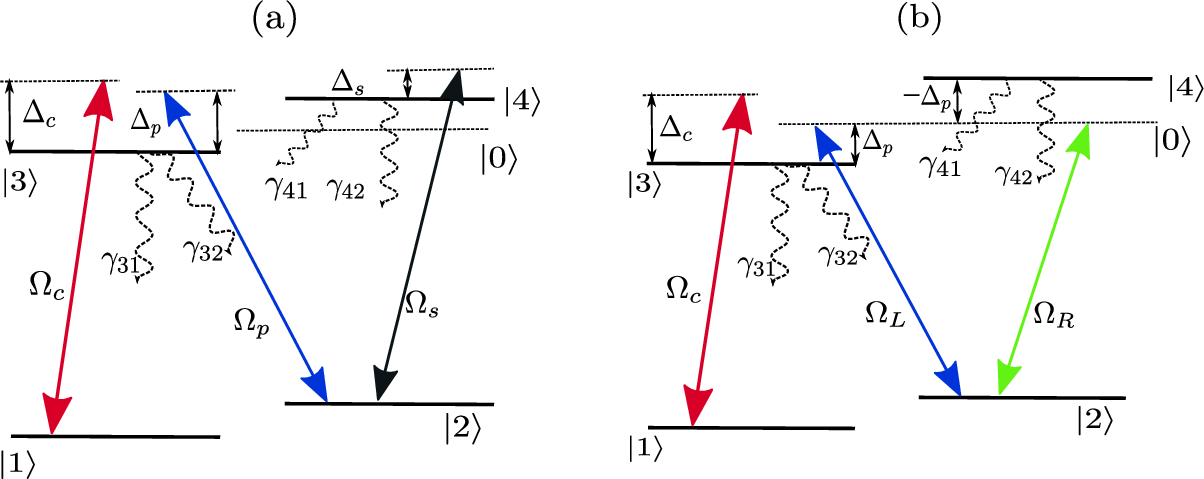}
	\caption{Schematic diagram of a four-level active-Raman-gain system. (a) Corresponds to the scalar probe connected to $|2\rangle\leftrightarrow|3\rangle$, and a signal field to $|2\rangle\leftrightarrow|4\rangle$ transition with a strong control field coupled to $|1\rangle\leftrightarrow|3\rangle$ transition. The probe, control, and signal detunings are denoted by $\Delta_p$, $\Delta_c$, and $\Delta_s$ respectively. (b) The probe beam is a VB. The right circularly polarized component, $E_R$, and the left circularly polarized component, $E_L$, of a weak probe VB drives the transitions, $|2\rangle\leftrightarrow|4\rangle$ and $|2\rangle\leftrightarrow|3\rangle$ respectively. The transition $|1\rangle\leftrightarrow|3\rangle$ is coupled by a strong control field $E_c$. The spontaneous emission decay rate from $|3\rangle$ and $|4\rangle$ states are given by $\gamma_{3j}$ and $\gamma_{4j}$ ($j\in 1, 2$).}
	\label{fig:1}
\end{figure}

\section{THEORETICAL FORMULATION}
\subsection{Level system}
In this paper, we conduct a systematic theoretical investigation of the focusing characteristics of scalar and VBs within a four-level ARG system. We demonstrate that both the linear and nonlinear properties of the probe beam significantly influence its propagation within the medium. Previous research has revealed intriguing phenomena associated with wave propagation in resonant optical media that feature an ARG core, supported by both theoretical and experimental findings  \cite{intro22,intro23,intro24}. Inspired by these investigations, this paper focuses on an ARG scheme involving two distinct configurations, utilizing a scalar probe beam and a VB, as illustrated in Figs. \ref{fig:1}(a) and \ref{fig:1}(b), respectively. We first address the scenario involving the scalar probe beam, followed by a discussion of the vector counterpart, concluding with a comparative analysis of the results from both configurations.

The implementation of the ARG scheme can be realized effectively in $^{87}$Rb $D_2$($5^2S_{1/2}\rightarrow5^2P_{3/2}$) transition hyperfine structure as: $|1\rangle = |5 ^2S_{1/2}, F = 1, m_F = 0 \rangle, |2\rangle = |5 ^2S_{1/2}, F = 2, m_F = 0 \rangle, |3\rangle = |5 ^2P_{3/2}, F = 1, m_F = -1 \rangle$, and $|4\rangle = |5 ^2P_{3/2}, F = 1, m_F = 1 \rangle$. The probe, control and the signal field is denoted as
\begin{align}
	&\Vec{E}_j(r,t) = \hat{e}_{j} \mathcal{E}_j(r) e^{-i(\omega_jt - k_jz)} + c.c.,
\end{align} 
where $\hat{e}_{j}$, $\mathcal{E}_j(r)$, $\omega_j$, $k_j$ are the polarization vector, spatial envelop, carrier frequency and wavevector respectively. The index $j\in p, c, s$ represents probe, control, and signal field respectively.

The time-dependent Hamiltonian describing the interaction of the model system as shown in Fig. \ref{fig:1}(a), can be written under dipole approximation as
\begin{subequations}
	\begin{align}
		\textbf{\textit{H}} =& \textbf{\textit{H}}_{0}+ \textit{\textbf{H}}_{I}, \\
		\textbf{\textit{H}}_{0} =&  \hbar (\omega_{21} |2 \rangle \langle 2 | + \omega_{31} |3 \rangle \langle 3 | + \omega_{41} |4 \rangle \langle 4 |) ,\\
		\textit{\textbf{H}}_{I} =& - \hat{d}.\vec{E}\nonumber\\
		=& -[\vec{d}_{42}.(\hat{e}_s \mathcal{E}_s e^{-i\omega_s t} + c.c.)|4\rangle \langle 2|\nonumber\\&+\vec{d}_{32}.(\hat{e}_p \mathcal{E}_p e^{-i\omega_p t} + c.c.)|3\rangle \langle 2|\nonumber\\&+\vec{d}_{31}.(\hat{e}_{c} \mathcal{E}_c e^{-i\omega_c t} + c.c.)|3\rangle \langle 1|] + \text{H.c.,}
	\end{align}
\end{subequations}
where $\omega_{j1}$($j = 2, 3, 4$) correspond to the frequency separation between the state $|j\rangle$ and the ground state $|1\rangle$ and $\vec{d}_{3k}$ ($k = 1, 2$), $\vec{d}_{42}$ are the matrix elements of the induced dipole moments for the transitions $|3\rangle\leftrightarrow |k\rangle$, and $|4\rangle\leftrightarrow|2\rangle$ respectively. In order to eliminate the explicit time dependence in the Hamiltonian, we perform the following unitary transformation
\begin{align}
	U =& \text{exp}[-i\omega_c t |3\rangle\langle 3| - i(\omega_c-\omega_p)t|2\rangle\langle 2| - i(\omega_s + \omega_c \nonumber\\&- \omega_p) t |4\rangle\langle 4|].
\end{align}
Now the effective Hamiltonian in the interaction picture is given as, $\mathcal{H} = \hat{U}^\dagger H \hat{U} - i\hbar\hat{U}^\dagger\partial_t \hat{U}$. Under the rotating wave approximation (RWA) it gives
\begin{align}
	\mathcal{H} =& -\hbar[(\Delta_c-\Delta_p)|2\rangle\langle 2|+\Delta_c|3\rangle\langle 3|+(\Delta_c + \Delta_s\nonumber\\& - \Delta_p)|4\rangle\langle 4|]-\hbar[\Omega_c|3\rangle\langle 1| +\Omega_p|3\rangle\langle 2|+\Omega_s|4\rangle\langle 2|] \nonumber\\& + \text{H.c.}
\end{align}
The single photon detunings of the probe and control field for their respective transitions are defined as
\begin{align}
	&\Delta_c = \omega_c - \omega_{31},\hspace{0.1cm} \Delta_p = \omega_p - \omega_{32},\hspace{0.1cm} \Delta_s = \omega_s - \omega_{42}, 
\end{align}
and the Rabi frequencies of probe field components and control field is written as
\begin{align}
	&\Omega_c = \dfrac{\vec{d}_{31}. \hat{e}_c}{\hbar}\mathcal{E}_c, \hspace{0.2cm}  \Omega_p = \dfrac{\vec{d}_{32}. \hat{e}_p}{\hbar}\mathcal{E}_p, \hspace{0.2cm}  \Omega_s = \dfrac{\vec{d}_{42}. \hat{e}_s}{\hbar}\mathcal{E}_s .
\end{align}
The dynamics of the system can be determined by the subsequent equations of motion \cite{intro25}
\begin{subequations}\label{eq.7}
	\begin{align}
		&i\dot{C}_{1} + \Omega_c^* C_3 = 0,\\
		&i\dot{C}_{2} + d_2 C_2 + \Omega_p^* C_3 + \Omega_s^* C_4 = 0,\\
		&i\dot{C}_{3} + d_3 C_2 + \Omega_c C_1 + \Omega_p C_2 = 0,\\
		&i\dot{C}_{4} + d_4 C_4 + \Omega_s C_2 = 0,
	\end{align} 
\end{subequations}
where, $\sum_{i=1}^{4}|C_{i}|^2=1$, $C_i (i\in 1$ $\text{to}$  4) is the probability amplitude of the bare state. In Eqs.(\ref{eq.7}a)-(\ref{eq.7}d), $d_2=(\Delta_c-\Delta_p+i\gamma_2),$ $d_3=(\Delta_c+i\gamma_3),$ $d_4=(\Delta_c+\Delta_s-\Delta_p+i\gamma_4)$ with $\gamma_j(j\in 2$ $\text{to}$ 4) being the atomic decay rates from the states $|j\rangle$. 
\subsection{Giant cross-Kerr nonlinearity}
In central symmetric materials, all even-order nonlinearities are identically equal to zero, resulting in the third-order nonlinearity being the lowest-order nonvanishing nonlinear optical susceptibility \cite{gk1}. Specifically, Kerr nonlinearity pertains to the third-order nonlinear optical susceptibility ($\chi^{(3)}$), which gives rise to the Kerr effect. The optical Kerr nonlinearity facilitates the propagation of ultrashort soliton-type pulses without temporal spreading. There is a strong preference for large third-order nonlinear susceptibilities under conditions of low optical power and high sensitivity. These properties can be effectively harnessed for the realization of single-photon nonlinear devices \cite{gk2,gk3}. To achieve this, it is essential for the linear susceptibilities to be minimized in comparison to the nonlinear susceptibilities. The Kerr effect manifests as an intensity-dependent alteration in the refractive index of a material, and is characterized by the real part of third order optical susceptibilities. The probe field susceptibility is precisely defined as
\begin{equation}\label{eq.8}
	\chi_{32}=\dfrac{\mathcal{N}|d_{23}|^2}{\hbar}\left(\dfrac{C_3 C_2^*}{\Omega_p}\right)\simeq \chi_p^{(1)} + \chi_{pp}^{(3)}|\Omega_p|^2 + \chi_{ps}^{(3)}|\Omega_s|^2,
\end{equation}
where, the linear probe susceptibility is denoted as $\chi_p^{(1)}$, while the third order self and cross Kerr susceptibilities are represented as $\chi_{pp}^{(3)}$ and $\chi_{ps}^{(3)}$ respectively. We solve Eqs. (\ref{eq.7}a)-(\ref{eq.7}d) under steady-state conditions. The probability amplitudes are obtained by satisfying the condition $\sum_{i=1}^{4}|C_{i}|^2=1$. Hence
\begin{equation}\label{eq.12}
	C_{1}=\biggl[1+|\Omega_c|^2 \dfrac{|D_s|^2 + |\Omega_p|^2\left(|\Omega_s|^2 + |d_4|^2\right)}{|D|^2}\biggr]^{-1/2},
\end{equation}	
where, $D_s = |\Omega_s|^2 - d_2 d_4$, and $D = d_4|\Omega_p|^2 + d_3|\Omega_s|^2 - d_2 d_3 d_4$. The other probability amplitudes can be written in terms of $C_1$ as
\begin{subequations}\label{eq.10}
	\begin{align}
		&C_{2} = -\dfrac{d_4 \Omega_p^* \Omega_c}{D}C_1,\\
		&C_{3} = -\dfrac{D_s \Omega_c}{D}C_1,\\
		&C_{4} = \dfrac{\Omega_p^* \Omega_c \Omega_s}{D}C_1.
	\end{align} 
\end{subequations}
In order to derive the linear and nonlinear susceptibilities as defined in Eq.  (\ref{eq.8}), a Taylor expansion is performed around $|\Omega_p|^2=|\Omega_s|^2=0$, resulting in the following expressions:
\begin{subequations}\label{eq.11}
	\begin{align}
		&\chi_{p}^{(1)} = -\dfrac{\mathcal{N}|d_{23}|^2}{\hbar}\dfrac{|\Omega_c|^2}{d_2^*\left(|d_3|^2+|\Omega_c|^2\right)},\\
		&\chi_{pp}^{(3)} = -\dfrac{\mathcal{N}|d_{23}|^2}{\hbar}\dfrac{|\Omega_c|^2\left(d_2^*d_3^*+d_2d_3-|\Omega_c|^2\right)}{d_2^*|d_2|^2\left(|d_3|^2+|\Omega_c|^2\right)^2},\\
		&\chi_{ps}^{(3)} = -\dfrac{\mathcal{N}|d_{23}|^2}{\hbar}\dfrac{|\Omega_c|^2 d_2d_4}{d_2^*|d_2|^2|d_4|^2\left(|d_3|^2+|\Omega_c|^2\right)}.
	\end{align} 
\end{subequations}
At the two photon resonance condition, $\textit{i.e.},$ $\Delta_{p}-\Delta_c$=0, we get
\begin{subequations}\label{eq.12}
	\begin{align}
		&\chi_{p}^{(1)} = -\dfrac{\mathcal{N}|d_{23}|^2}{\hbar}\dfrac{i |\Omega_c|^2}{\gamma_2\left(\Delta_c^2+\gamma_3^2+|\Omega_c|^2\right)},\\
		&\chi_{pp}^{(3)} = \dfrac{\mathcal{N}|d_{23}|^2}{\hbar}\dfrac{i|\Omega_c|^2\left(2\gamma_2\gamma_3+|\Omega_c|^2\right)}{\gamma_2^3\left(\Delta_c^2+\gamma_3^2+|\Omega_c|^2\right)^2},\\
		&\chi_{ps}^{(3)} = \dfrac{\mathcal{N}|d_{23}|^2}{\hbar}\dfrac{|\Omega_c|^2\left(\Delta_s+i\gamma_4\right)}{\gamma_2^2\left(\Delta_s^2+\gamma_4^2\right)\left(\Delta_c^2+\gamma_3^2+|\Omega_c|^2\right)}.
	\end{align} 
\end{subequations}
From Eqs. (\ref{eq.12}a)-(\ref{eq.12}c), it is observed that $\chi_{p}^{(1)}$ and $\chi_{pp}^{(3)}$ are entirely imaginary. The negative sign associated with Im[$\chi_{p}^{(1)}$] indicates linear gain, while Im[$\chi_{pp}^{(3)}$] reflects nonlinear absorption. This nonlinear absorption, arising from the self-Kerr effect, increases with the intensification of the probe field intensity. Conversely, $\chi_{ps}^{(3)}$ comprises both real and imaginary components. Notably, only Re[$\chi_{ps}^{(3)}$] at two-photon resonance results in significant cross-Kerr nonlinearity. Moreover, Im[$\chi_{ps}^{(3)}$] characterizes the nonlinear absorption induced by the cross-Kerr effect, which also escalates with an increase in the intensity of the signal field.

\subsection{Beam propagation equation with paraxial approximation}
The investigation of beam propagation equations plays a pivotal role in examining the influence of absorption, diffraction, dispersion on probe beam propagation. By employing the slowly varying envelope and paraxial wave approximations, the propagation equations governing the probe beam can be written as

(i) $\textit{For scalar probe, and control beam:}$
\begin{subequations}{\label{eq.13}}
	\begin{align}
		&\dfrac{\partial\Omega_p}{\partial z} = \dfrac{i}{2k_p}\nabla^2_{\perp}\Omega_p+ 2\pi i k_p \chi_{32} \Omega_p,\\
		&\dfrac{\partial\Omega_c}{\partial z} = \dfrac{i}{2k_c}\nabla^2_{\perp}\Omega_c+ 2\pi i k_c \chi_{31} \Omega_c,
	\end{align}
\end{subequations}
and

(ii) $\textit{For vector probe beam:}$
\begin{subequations}{\label{eq.14}}
	\begin{align}
		&\dfrac{\partial\Omega_R}{\partial z} = \dfrac{i}{2k_R}\nabla^2_{\perp}\Omega_R+ 2\pi i k_R \chi_{42} \Omega_R,\\
		&\dfrac{\partial\Omega_L}{\partial z} = \dfrac{i}{2k_L}\nabla^2_{\perp}\Omega_L+ 2\pi i k_L \chi_{32} \Omega_L.
	\end{align}
\end{subequations}
Note that Eqs. (\ref{eq.13}a), and (\ref{eq.13}b) refer to scalar beams and Eqs. (\ref{eq.14}a), and (\ref{eq.14}b) correspond to VB propagation. The right-hand side of the equations consists of two terms: the first term accounts for diffraction, while the second term denotes the dispersion and absorption of the probe, through its susceptibility. Throughout our work, we consistently operate within the paraxial regime, a framework that is appropriate given that the beam waist is considerably larger than the wavelength of the light used. It is crucial to consider the dynamics of the strong control field, as it is associated with both the ground state and the excited state, thus impacting the propagation of the probe beam. Given that the signal field $E_s$ is a strong plane wave connected with $|2\rangle\leftrightarrow|4\rangle$ transition, its propagation effects are negligible and have been validated through detailed simulation analysis. To numerically analyze Eqs. (\ref{eq.13}a), (\ref{eq.13}b), and Eqs. (\ref{eq.14}a), (\ref{eq.14}b), the split-step Fourier method (SSFM) has been selected for the study.
\begin{figure}[ht]
	\centering
	\includegraphics[width=\linewidth]{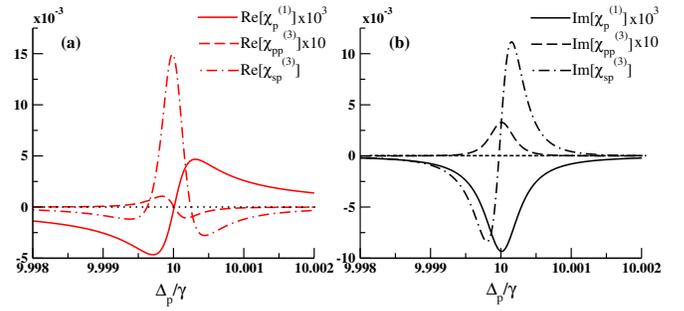}
	\caption{Real and imaginary part of both linear and third-order nonlinear $\chi_{32}$ are plotted against the probe detuning. (a), (b) correspond to real, and imaginary part of $\chi_{32}$ respectively. Parameter used: $\Omega_{p}=0.001\gamma,\Omega_{c}=0.7\gamma$, $\Omega_s = 0.05\gamma$. $\gamma_{2} = 3\times10^{-4}\gamma, \gamma_{3} = 5\times10^2\gamma$, $\gamma_{4} = 0.5\gamma$. $\Delta_c = 10\gamma$,$\Delta_s = 2\gamma$. Normalized frequency $\gamma = 10^6$ $\text{s}^{-1}$. The density of atoms, $\mathcal{N} = 5\times10^{11}\text{cm}^{-3}$. }
	\label{fig:2}
\end{figure}
\section{NUMERICAL RESULTS}
\subsection{Linear and nonlinear susceptibilities}
In this section, we delve into the analysis of both linear and nonlinear susceptibilities observed in the probe beam. Figures \ref{fig:2} (a) and \ref{fig:2}(b) effectively illustrate the behavior of these susceptibilities by showcasing the real and imaginary components of both the linear susceptibility and the third order Kerr nonlinear susceptibility. In Fig. \ref{fig:2} (a), we can see that the 
\begin{figure*}
	\centering
	\includegraphics[width=\linewidth]{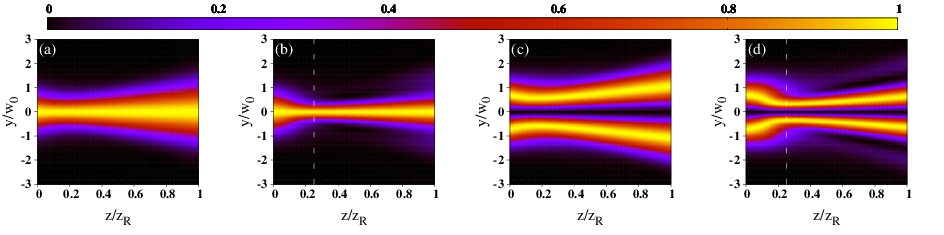}
	\caption{Longitudinal intensity profile of (a) Gaussian, and (c) LG beam propagating inside the medium without taking the effect of cross-Kerr nonlinearity. (b), and (d) corresponds to including the effect of cross-Kerr nonlinearity respectively. The value of $l_c$ for the propagation of probe beam is taken as 1. The probe and the control beam have a waist, $w_0 = 50\mu m$ and is maintained consistently. Other parameters are same as Fig. \ref{fig:2}.}
	\label{fig:3}
\end{figure*}
real part of the linear susceptibility, denoted as Re[$\chi_{p}^{(1)}$], along with the real part of the third order susceptibility, Re[$\chi_{pp}^{(3)}$], both approach zero at the point of two-photon resonance. In contrast, Re[$\chi_{sp}^{(3)}$] exhibits a significant nonzero value, indicating a notable response under these conditions. This distinct behavior suggests that while the linear susceptibility is minimal at resonance, the cross-Kerr nonlinear response is greatly enhanced. In Fig. \ref{fig:2}(b), it is observed that Im[$\chi_{p}^{(1)}$] exhibits a negative value at the two-photon resonance, which is indicative of linear gain. The strong control field interacts with the ground state $|1\rangle$ and the excited state $|3\rangle$, facilitating a significant population transfer from the initially populated ground state $|1\rangle$ to the excited state $|3\rangle$. This transfer of population leads to gain in the $|1\rangle \rightarrow |3\rangle$ transition. Consequently, the gain in the probe field can be modulated by appropriately selecting the one-photon detuning of the control field. Additionally, the continuous wave (CW) signal field, which is coupled to the $|2\rangle \leftrightarrow |4\rangle$ transition, contributes to the cross-Kerr nonlinearity of the system. It is important to note that at two-photon resonance, both Im[$\chi_{pp}^{(3)}$] and Im[$\chi_{sp}^{(3)}$] are positive, indicating the presence of nonlinear absorption in the probe field.
\subsection{Beam propagation inside medium}
In order to achieve the focusing of the probe beam, it is crucial to meticulously select the transverse spatial profile of the control beam. This selection involves ensuring that the intensity at the center of the control beam significantly exceeds that of its outer regions. For this
purpose, we have opted for a Gaussian-shaped control beam. This choice is made to effectively support the propagation of both Gaussian and Laguerre-Gaussian (LG) probe beams. The control field is defined as follows:
\begin{equation}
	\Omega_{c}(r,\phi,0) = \Omega_{c0} \text{exp}\biggl[-\left(\dfrac{r^2}{w_0^2}\right)^{l_c}\biggr].
\end{equation}
The input amplitude of the control beam is represented by $\Omega_{c0}$, and $l_c$ is a positive integer. The value $l_c \ne 1$ corresponds to a super-Gaussian or flat-top beams. In Figs. \ref{fig:3}(b) and \ref{fig:3}(d), we provide a detailed illustration of the behavior of Gaussian and LG probe beams as they propagate through the medium, taking into account the effects of cross-Kerr nonlinearity. For comparative purposes, we have also depicted the beam propagation in the absence of cross-Kerr nonlinearity, as illustrated in Figs. \ref{fig:3}(a) and \ref{fig:3}(c). In Figs. \ref{fig:3}(a) and \ref{fig:3}(c), it is evident that the probe beams undergo significant diffraction as they propagate through the medium. Initially, the probe beam converges towards the center due to rapid energy transfer from the control beam, resulting in a gain-narrowed region. However, subsequent to this process, the medium is unable to counterbalance the diffraction, as both Re[$\chi_{p}^{(1)}$] and Re[$\chi_{pp}^{(3)}$] are zero as depicted in Fig. \ref{fig:2}(a). In contrast, an examination of Fig. \ref{fig:2}(a) reveals that Re[$\chi_{sp}^{(3)}$] displays a significant cross-Kerr nonlinearity at resonance. The introduction of a three-photon off-resonance condition, specifically $\Delta_s \neq 0$, is essential for attaining a marked cross-Kerr effect. This noteworthy nonlinearity serves to mitigate the diffraction, thereby facilitating the focusing of the probe beams, as illustrated in Figs. \ref{fig:3}(b) and \ref{fig:3}(d). However, as the beam converges, diffraction gradually increases due to the positive value of Im\([\chi_{pp}^{(3)}]\), which has a defocusing effect. Consequently, the contrasting effects of the presence and absence of cross-Kerr interactions are prominently emphasized, highlighting their essential significance in beam manipulation within the medium.
\section{Comparative study with probe VB}
Till now, we have examined the model presented in Fig. \ref{fig:1}(a), in which the scalar beam is coupled to the transition between states $|2\rangle$ and $|3\rangle$. An alternative perspective is now considered, wherein a VB serves as a weak probe beam, as illustrated in Fig. \ref{fig:1}(b). The two orthogonal components of the VB are associated with the transitions $|2\rangle \leftrightarrow |3\rangle$ and $|2\rangle \leftrightarrow |4\rangle$, corresponding to left and right circular polarizations, respectively. Unlike scalar beams, VBs exhibit an inhomogeneous polarization distribution across the transverse plane. This polarization distribution can be classified into two categories. A cylindrical vector (CV) beam is characterized by radial, azimuthal, and spiral profiles with respect to the axial symmetry of the beam. Conversely, a full Poincaré (FP) beam displays a transverse polarization distribution that varies in both azimuthal and radial directions, resulting in configurations such as lemon, star, and web.

The electric field associated with the probe VB can be expressed as
\begin{equation}\label{eq.15}
	\vec{E}(r,\phi,z) = \mathcal{E}_L(r,\phi,z)\hat{e}_L + \mathcal{E}_R(r,\phi,z)\hat{e}_R,
\end{equation}
where the left circularly polarized component $
\mathcal{E}_L = \text{cos}(\alpha)LG_0^{l_L}$, and right circularly polarized component $\mathcal{E}_R = \text{sin}(\alpha)e^{i\theta}LG_0^{l_R}$. $LG_0^{l_i}(i = L, R)$, refer to the Laguerre Gaussian polynomial, having radial index zero and is given by 
\begin{align}\label{eq.17}
	LG_0^{l_i}(r, \phi, z) =& \mathcal{E}_0 \sqrt{\dfrac{2}{\pi|l_i|!}} \left(\dfrac{r\sqrt{2}}{w(z)}\right)^{|l_i|}e^{\scalebox{0.8}{$-\dfrac{r^2}{w(z)^2}$}}e^{il_i\phi+ik_in_iz}\nonumber \\ &\times  \text{exp}\left(\dfrac{ik_in_ir^2z}{2(z^2+n_i^2z_R^2)}\right)e^{-i(|l_i| + 1)\eta(z)}.
\end{align}
The beam radius at a propagation length $z$ is $w(z) = w_0\sqrt{1+z^2/n_i^2z_R^2}$ , where $w_0$ represents the beam waist at $z = 0$, and $n_i$ is the refractive index of the medium. $z_R = k_i w_0^2/2$ denotes the free space Rayleigh length, with $k_i$ being the free space wave number. The OAM index is $l_i$, and $(|l_i| + 1)\eta(z)$ is the Gouy phase, where $\eta(z) = \text{tan}^{-1}(z/n_iz_R)$. Both the refractive indices, $n_R = 1+2\pi \text{Re}[\chi_{42}]$, and $n_L = 1+2\pi \text{Re}[\chi_{32}]$. To fully characterize the state of polarization of light, it is essential to utilize four Stokes parameters. These parameters are defined as follows \cite{com1}:
\begin{align}\label{eq.18}
	&S_0 = |\mathcal{E}_R|^2 + |\mathcal{E}_L|^2,\hspace{0.2cm} S_1 = 2\text{Re}[\mathcal{E}_R^* \mathcal{E}_L],\nonumber\\ 
	&S_2 = 2\text{Im}[\mathcal{E}_R^* \mathcal{E}_L],\hspace{0.2cm} S_3 = |\mathcal{E}_L|^2 - |\mathcal{E}_R|^2.
\end{align}
The ellipticity, denoted as $\zeta$, and the orientation of polarization, represented by $\xi$, at each point within the transverse plane are defined as follows
\begin{align}\label{eq.19}
	&\zeta = \dfrac{1}{2} \text{sin}^{-1}\left(\dfrac{S_3}{S_0}\right), \hspace{0.2cm} \xi = \dfrac{1}{2} \text{tan}^{-1}\left(\dfrac{S_2}{S_1}\right).
\end{align} 
Assuming that the free space wave vector $k_R = k_L = k$, gives
\begin{align}\label{eq.20}
	\xi(z) =& -\dfrac{1}{2}\biggl[\theta + \phi\Delta(l_{L,R}) + kz\Delta(n_{R,L})+\eta(z)\Delta(|l_{L,R}|)\nonumber \\&+
	\dfrac{kzr^2}{2}\biggl\{\dfrac{n_R}{z^2+n_R^2z_R^2}-\dfrac{n_L}{z^2+n_L^2z_R^2}\biggl\}\biggl],
\end{align}
where $\Delta(l_{L,R}) = l_L-l_R$, $\Delta(|l_{L,R}|) = |l_L|-|l_R|$, and $\Delta(n_{R,L}) = n_R-n_L$. Following propagation through a distance $z$ within the medium, the polarization of a VB at each point on the transverse plane experiences a rotation described by the following expression:
\begin{align}\label{eq.21}
	\Delta\xi(z) =& -\dfrac{\Delta|\l_{L,R}|\eta(z)}{2}-\dfrac{1}{2}\biggl[
	\dfrac{kzr^2}{2}\biggl\{\dfrac{n_R}{z^2+n_R^2z_R^2}\nonumber \\&-\dfrac{n_L}{z^2+n_L^2z_R^2}\biggl\}+kz\Delta(n_{R,L})\biggl].
\end{align}
According to Eq. (\ref{eq.21}), it is clear that polarization rotation within the medium for CV beams occurs solely due to the difference in the refractive indices of the two components of the probe beam as $|l_L| = |l_R|$. Furthermore, the change in the ellipticity of the VB beam can also be described as
\begin{align}\label{eq.22}
	\Delta\zeta(z) =& \dfrac{1}{2}\biggl[\text{sin}^{-1}\biggl\{
	\dfrac{1 - a \text{tan}^2\alpha}{1 + a \text{tan}^2\alpha}\biggl\}- \text{sin}^{-1}\biggl\{\dfrac{1 -  b\text{tan}^2\alpha}{1 +  b\text{tan}^2\alpha}\biggl\}\biggl],
\end{align}
where

\begin{align}\label{eq.23}
	a=&\text{exp}\biggl[-\dfrac{2r^2}{w_0^2}\left(\dfrac{n_R^2z_R^2}{z^2+n_R^2z_R^2}-\dfrac{n_L^2z_R^2}{z^2+n_L^2z_R^2}\right)\biggl]\times\nonumber \\ &\biggl\{\dfrac{r\sqrt{2}}{w_0\sqrt{1+z^2/n_R^2z_R^2}}\biggl\}^{2|l_R|} \biggl\{\dfrac{r\sqrt{2}}{w_0\sqrt{1+z^2/n_L^2z_R^2}}\biggl\}^{-2|l_L|},
\end{align}

and
\begin{align}\label{eq.24}
	b=&\left(\dfrac{r\sqrt{2}}{w_0}\right)^{2\left(|l_R|-|l_L|\right)}.
\end{align} 
\subsection{Suceptibilities of the probe VB}
The VB constitutes a superposition of two orthogonal components, leading to distinct susceptibilities for the probe VB. These susceptibilities for each component can be defined as follows:
\begin{subequations}\label{eq.25}
	\begin{align}
		\chi_{32}\equiv\chi_L=\dfrac{\mathcal{N}|d_{23}|^2}{\hbar}\left(\dfrac{C_3 C_2^*}{\Omega_L}\right)\simeq & \chi_L^{(1)} + \chi_{LL}^{(3)}|\Omega_L|^2\nonumber \\& + \chi_{LR}^{(3)}|\Omega_R|^2,\\
		\chi_{42}\equiv\chi_R=\dfrac{\mathcal{N}|d_{24}|^2}{\hbar}\left(\dfrac{C_4 C_2^*}{\Omega_R}\right)\simeq & \chi_R^{(1)} + \chi_{RR}^{(3)}|\Omega_R|^2 \nonumber \\ &+ \chi_{RL}^{(3)}|\Omega_L|^2.
	\end{align}
\end{subequations}
The linear probe susceptibility is denoted as $\chi_{L,R}^{(1)}$, while the third order self and cross Kerr susceptibilities are represented as $\chi_{LL,RR}^{(3)}$ and $\chi_{LR,RL}^{(3)}$ respectively. As before we perform a Taylor expansion around $|\Omega_L|^2=|\Omega_R|^2=0$, resulting in the following expressions:
\begin{subequations}\label{eq.26}
	\begin{align}
		&\chi_{L}^{(1)} = -\dfrac{\mathcal{N}|d_{23}|^2}{\hbar}\dfrac{|\Omega_c|^2}{d_2^*\left(|d_3|^2+|\Omega_c|^2\right)},\\
		&\chi_{LL}^{(3)} = -\dfrac{\mathcal{N}|d_{23}|^2}{\hbar}\dfrac{|\Omega_c|^2\left(d_2^*d_3^*+d_2d_3-|\Omega_c|^2\right)}{d_2^*|d_2|^2\left(|d_3|^2+|\Omega_c|^2\right)^2},\\
		&\chi_{LR}^{(3)} = -\dfrac{\mathcal{N}|d_{23}|^2}{\hbar}\dfrac{|\Omega_c|^2 d_2d'_4}{d_2^*|d_2|^2|d'_4|^2\left(|d_3|^2+|\Omega_c|^2\right)},\\
		&\chi_{R}^{(1)} = \chi_{RR}^{(3)} = 0,\\
		&\chi_{RL}^{(3)} = -\dfrac{\mathcal{N}|d_{24}|^2}{\hbar}\dfrac{|\Omega_c|^2}{|d_2|^2 d'_4\left(|d_3|^2+|\Omega_c|^2\right)}.
	\end{align} 
	
\end{subequations} 
\begin{figure}[hb]
	\centering
	\includegraphics[width=\linewidth]{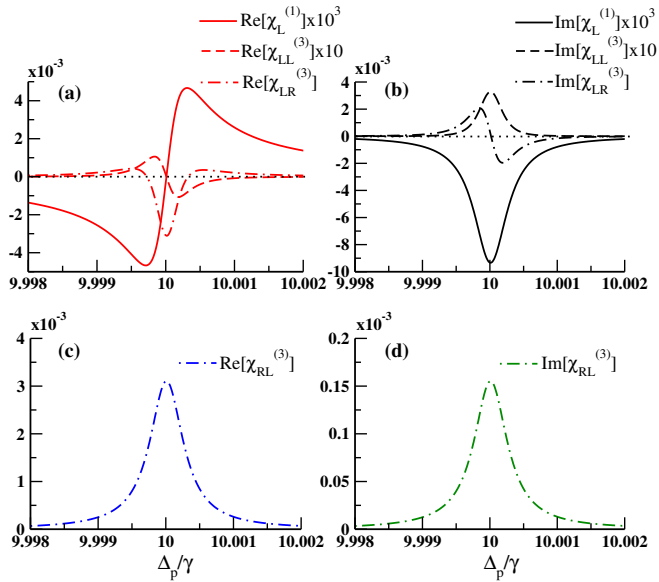}
	\caption{Real and imaginary part of both linear and third-order nonlinear $\chi_{L}$ and $\chi_{R}$ are plotted against the corresponding transition detunings. (a), (b) correspond to real, and imaginary part of $\chi_L$ plotted against $\Delta_p/\gamma$ respectively. Similarly (c), (d) correspond to real, and imaginary part of $\chi_R$ plotted against $\Delta_p/\gamma$ respectively. Parameter used: $\Omega_{R,L}=0.001\gamma,\Omega_{c}=0.7\gamma$. Other parameters are same as Fig. \ref{fig:2} }
	\label{fig:4}
\end{figure} 

At the two photon resonance condition, we get
\begin{subequations}\label{eq.27}
	\begin{align}
		&\chi_{L}^{(1)} = -\dfrac{\mathcal{N}|d_{23}|^2}{\hbar}\dfrac{i |\Omega_c|^2}{\gamma_2\left(\Delta_c^2+\gamma_3^2+|\Omega_c|^2\right)},\\
		&\chi_{LL}^{(3)} = \dfrac{\mathcal{N}|d_{23}|^2}{\hbar}\dfrac{i|\Omega_c|^2\left(2\gamma_2\gamma_3+|\Omega_c|^2\right)}{\gamma_2^3\left(\Delta_c^2+\gamma_3^2+|\Omega_c|^2\right)^2},\\
		&\chi_{LR}^{(3)} = \dfrac{\mathcal{N}|d_{23}|^2}{\hbar}\dfrac{|\Omega_c|^2\left(-\Delta_p+i\gamma_4\right)}{\gamma_2^2\left(\Delta_p^2+\gamma_4^2\right)\left(\Delta_c^2+\gamma_3^2+|\Omega_c|^2\right)},\\
		&\chi_{R}^{(1)} = \chi_{RR}^{(3)} = 0,\\
		&\chi_{RL}^{(3)} = \dfrac{\mathcal{N}|d_{23}|^2}{\hbar}\dfrac{ |\Omega_c|^2\left(\Delta_p+i\gamma_4\right)}{\gamma_2^2\left(\Delta_p^2+\gamma_4^2\right)\left(\Delta_c^2+\gamma_3^2+|\Omega_c|^2\right)},
	\end{align} 
\end{subequations}
where, $d'_4=(\Delta_c-2\Delta_p+i\gamma_4)$.
In the preceding equations, it is interesting to note that the right component of the probe VB, connected to the transition $|2\rangle\leftrightarrow|4\rangle$, exhibits only cross-Kerr susceptibility.

From Eqs. (\ref{eq.27}a)-(\ref{eq.27}e), it can be observed that $\chi_{L}^{(1)}$ and $\chi_{LL}^{(3)}$ are entirely imaginary quantities. The negative sign associated with Im[$\chi_{L}^{(1)}$] indicates a linear gain, while Im[$\chi_{LL}^{(3)}$] reflects nonlinear absorption phenomena. In contrast, $\chi_{LR}^{(3)}$ exhibits both real and imaginary components. It is determined that the left circular component of the probe beam contributes to both linear gain and nonlinear absorption, with nonlinear absorption increasing in response to higher probe beam intensities. For the right circular component, both the linear and self-Kerr terms are found to be zero. This phenomenon arises because the coherence in the transition $|4\rangle \leftrightarrow |2\rangle$ develops solely in the presence of the left circular component. However, there is a contribution from the term $\chi_{RL}^{(3)}$, which results in absorption due to cross-Kerr nonlinearity. In Figs. \ref{fig:4}(a)-(d) we plot the linear and nonlinear responses of both the components of the probe beam which illustrate the above facts. It is noteworthy that the two components of the VB are not separable, leading to an enhancement of nonlinear interactions as the probe beam intensity increases.
\begin{figure*}
	\centering
	\includegraphics[width=\linewidth]{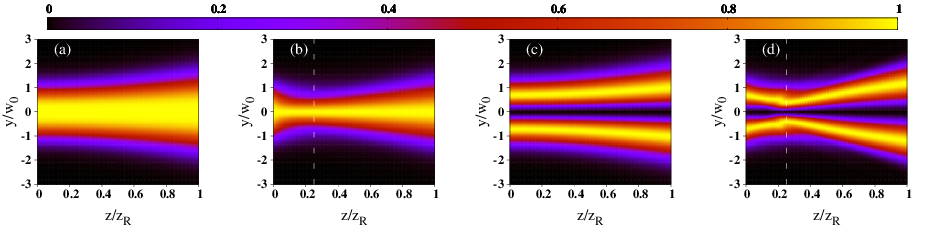}
	\caption{Longitudinal intensity profile of (a) lemon ($l_L = 0, l_R = 1, \alpha = \pi/4, \theta = 0$), (c) radial ($l_L = -1, l_R = 1, \alpha = \pi/8, \theta = 0$) VB propagating in free space, and inside the gain medium (b), and (d) respectively. The value of $l_c$ for the propagation of lemon and radial VB is taken as 1, and 4 respectively. The probe and the control beam have a waist, $w_0 = 50\mu m$ and is maintained consistently. White dashed vertical line denotes the minimum spot size achieved due to gain-induced narrowing. Other parameters are same as Fig. \ref{fig:4}.}
	\label{fig:5}
\end{figure*}
\subsection{Gain induced narrowing} 
\begin{figure}[hb]
	\centering
	\includegraphics[width=\linewidth]{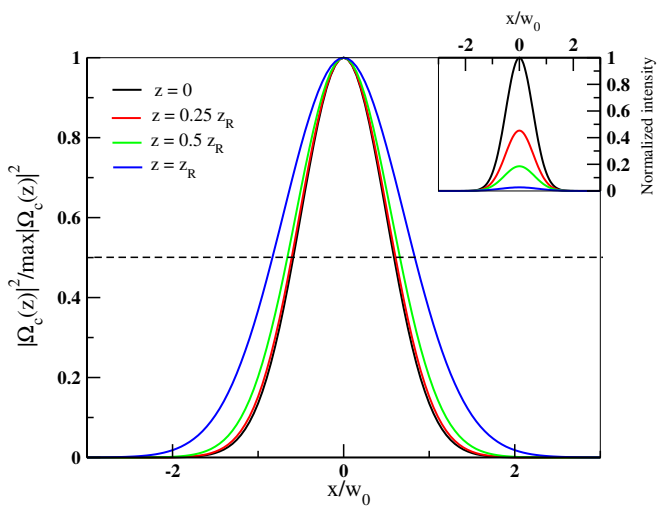}
	\caption{The beam waist of the control field is measured at different propagation distances. The variation of its intensity is presented in the inset.}
	\label{fig:6}
\end{figure}

In this section we demostrate the VB propagation inside the atomic medium. In Figs. \ref{fig:5}(a)-(d), we illustrate the propagation of VBs in both free space and within the Raman gain medium. During the initial narrowing phase, the control beam facilitates energy transfer to the probe beam. The energy transfer is more pronounced in regions with higher control beam intensity, resulting in a gain-narrowed probe profile that achieves a minimum spot radius. In the illustrations labeled Fig. \ref{fig:5}(a) and Fig. \ref{fig:5}(c), the longitudinal profile of lemon and radial VB propagation in free space is depicted. Conversely, Fig. \ref{fig:5}(b) and Fig. \ref{fig:5}(d) characterize the propagation of lemon and radial VB inside the medium. Although the consequence of cross-Kerr nonlinearity is not evident here, the linear gain of the system supports the narrowing of the VBs. This phenomenon is known as gain-induced narrowing. Refractive focusing is attributed to the real part, while gain narrowing is linked to the imaginary part of the susceptibility of the probe beam. The control beam has a significant influence on the propagation of the probe beam. In Fig. \ref{fig:6} we have plotted the profile of the control beam that affects the propagation of the lemon VB at various distances. The beam width at the full width at half maximum (FWHM) for propagation distances of $z$ = 0, 0.25$z_R$, 0.5$z_R$, and $z_R$ is measured to be 1.16$w_0$, 1.22$w_0$, 1.32$w_0$, and 1.66$w_0$, respectively. In the inset of Fig. \ref{fig:6}, we present the normalized intensity of the control beam at the specified distances. It is evident that the gain-assisted narrowing of the probe beam results from the transfer of energy from the control beam to the probe beam. Consequently, as the energy transfer diminishes, the probe beam begins to diverge. These observations highlight the necessity of carefully controlling the intensities of both the probe and control beams in order to achieve a moderate gain regime, which in turn facilitates the narrowing of VBs. The discussed gain-induced narrowing
phenomenon is expected to be a pervasive occurrence, applicable to any medium supporting nonlinear focusing and stimulated Raman scattering.

\begin{figure}[hb]
	\centering
	\includegraphics[width=\linewidth]{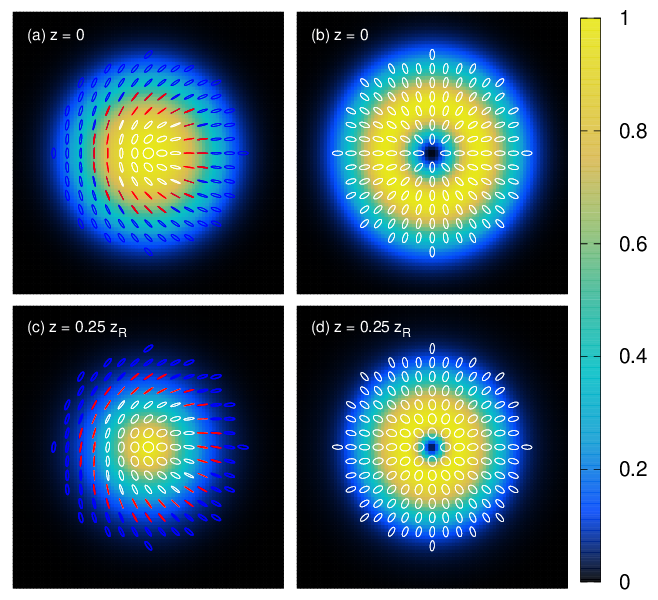}
	\caption{Transverse intensity and polarization distribution at the $z=0$ for (a) lemon, (b) radial and at minimum beam radius $z=0.25z_R$ for (c) lemon, (d) radial VB respectively. The intensity is normalized by its maximum intensity for each VB. The colors white, red, and blue correspond to left circular, linear, and right circular polarizations, respectively. Parameters remain same as Fig. \ref{fig:5}.}
	\label{fig:7}
\end{figure}

Further, we examine the state of polarization (SOP) of the probe VBs as it poses inhomogeneous polarization distribution unlike scalar beam. As depicted in Fig. \ref{fig:7}(a) and 
Fig. \ref{fig:7}(b), we display the SOP at input $z = 0$ for the lemon and radial VBs, respectively. Subsequently, after the beam narrowing, the SOP of the lemon and radial VBs at $z = 0.25z_R$ is illustrated by Fig. \ref{fig:7}(c) and Fig. \ref{fig:7}(d) correspondingly. Our findings indicate that the SOP of the lemon VB at $z = 0.25z_R$ has undergone a clockwise rotation by an angle of $\pi/4$ with a variation in ellipticity. The disparity between the refractive indices $n_R$ and $n_L$ is central to this phenomenon. In contrast, the radial VB exhibits an ellipticity mostly unchanged. Moreover, we observe that the FWHM of the lemon VB at the input is 1.84$w_0$, which subsequently becomes 0.98$w_0$ upon narrowing at z = 0.25$z_R$. Likewise, at the input, the FWHM of the central dark region of the radial VB is 0.68$w_0$, reduced to 0.36$w_0$ upon narrowing. Furthermore, as the VB propagates and diverges within the medium, its SOP gradually becomes more homogeneous across the transverse plane, resembling the behavior of a scalar beam. This phenomenon becomes evident when there exists an insufficient difference between the refractive indexes $n_R$ and $n_L$.  
\begin{table}[t]
	\caption{\label{tab:table1}%
		Comparison of linear and nonlinear susceptibilities at two photon resonance, type of beam focusing or narrowing and spot size reduction between the scalar (Fig. \ref{fig:1}(a)), and vector (Fig. \ref{fig:1}(b)) probe beam in ARG scheme.
	}
	\begin{ruledtabular}
		\begin{tabular}{llll}
			\textrm{}&
			\textrm{Scalar probe}&
			\textrm{}&
			\textrm{Vector probe}\\
			\colrule
			Re[$\chi_p^{(1)}$] & 0 & Re[$\chi_L^{(1)}$]& 0 \\
			Re[$\chi_{pp}^{(3)}$] & 0 & Re[$\chi_{LL}^{(3)}$] & 0\\
			Re[$\chi_{ps}^{(3)}$] & 14.6 & Re[$\chi_{LR}^{(3)}$] & -3.11 \\
			&  & Re[$\chi_{RL}^{(3)}$] & 3.11 \\
			\colrule 
			Im[$\chi_p^{(1)}$] & -9.34$\times 10^{-3}$ & Im[$\chi_L^{(1)}$]& -9.34$\times 10^{-3}$  \\
			Im[$\chi_{pp}^{(3)}$] & 0.328 & Im[$\chi_{LL}^{(3)}$] & 0.328\\
			Im[$\chi_{ps}^{(3)}$] & 3.66 & Im[$\chi_{LR}^{(3)}$] & 0.16 \\
			&  & Im[$\chi_{RL}^{(3)}$] & 0.16 \\
			\colrule 
			Focusing & Cross-kerr induced & Narrowing & Gain-induced \\
			\colrule 
			Spot size & 56\% (for gaussian) & Spot size & 46.7\% (for Lemon) \\
			reduction & 53\% (for LG) & reduction & 47\% (for radial)\\
			
		\end{tabular}
	\end{ruledtabular}
\end{table}
Finally, we present a concise overview of the optical properties associated with two configurations employing scalar and vector probe beams. In Table \ref{tab:table1}, we provide the values of the real and imaginary components of linear and nonlinear susceptibilities under two-photon resonance conditions, along with details regarding the nature of focusing or narrowing, and the reduction in spot size at the minimum beam waist for both scenarios. It is observed that the scalar probe beam exhibits cross-Kerr-induced focusing due to the significant magnitude of the real part of its cross-Kerr susceptibility. Conversely, in the case of the vector probe beam, the real part of the cross-Kerr susceptibility for the left and right circular components is equivalent in magnitude but opposite in sign. Notably, the nonlinear absorption for the vector probe beam is considerably lower than that of the scalar probe beam. However, the linear gain remains consistent for both configurations. Consequently, gain-assisted narrowing is discerned in the VB scenario.

\section{CONCLUSION}
In conclusion, this study conducts a theoretical examination of the focusing characteristics of scalar and VBs within the framework of a four-level active Raman gain system. In both configurations utilizing scalar and vector probe beams, the ground state is coupled with a strong control field characterized by high detuning, which is indicative of a gain system. The CW signal field, with a nonzero detuning, serves to enhance the cross-Kerr nonlinearity of the system. With a precise selection of input field intensities and the spatial profile of the control beam, both scalar and vector probe beams exhibit a focusing or narrowing phenomena. However, it is observed that the medium's susceptibility varies distinctly between the scalar and vector probe beams. While the combination of linear and self-Kerr nonlinearity can focus the scalar probe beam to a certain degree, the impact of cross-Kerr nonlinearity allows for focusing of the beam into a significantly smaller spot. Conversely, the vector probe beam showcases gain-assisted narrowing as a result of energy transfer from the control beam. Furthermore, an analysis of the SOP of the VB is conducted at the minimum beam waist. The focusing or the narrowing effects observed in both scalar and VBs lead to a reduction in spot size, which has the potential to substantially enhance optical resolution.
\begin{acknowledgments}
We gratefully acknowledge funding by the Department of Science and Technology, Anusandhan National Research Foundation, Government of India (Grant No. CRG/2023/001318).
\end{acknowledgments}
\twocolumngrid
\bibliography{paper}
\end{document}